\documentclass[12pt]{article}
\usepackage{epsfig}
\def\beq{\begin{equation}}
\def\eeq#1{\label{#1}\end{equation}}
\def\eeqn{\end{equation}}

\def\beqa{\begin{eqnarray}}
\def\eeqa#1{\label{#1}\end{eqnarray}}
\def\eeqan{\end{eqnarray}}

\let\bar=\overbar

\def\Dslash{\not{\hbox{\kern-4pt $D$}}}
\def\dslash{\not{\hbox{\kern-2pt $\del$}}}

\def\ee{e^+e^-}

\def\msb{{\bar{\ssstyle M \kern -1pt S}}}

\def\pz{\pi^0}

\def\ll{l\bar l}
\def\ee{e^+e^-}
\def\ep{\epsilon}
\def\mm{\mu^+\mu^-}
\def\g{\gamma}
\def\Title#1{\begin{center} {\Large {\bf #1} } \end{center}}

\begin{document}

\Title{KTeV Results on Chiral Perturbation Theory}

\begin{center}{\large \bf Contribution to the proceedings of HQL06,\\
Munich, October 16th-20th 2006}\end{center}

\bigskip\bigskip

\begin{raggedright}  

{\it Elliott Cheu\index{Cheu, E.}\\
Department of Physics\\
1118 E 4th Street\\
Tucson, AZ 85749 USA}
\bigskip\bigskip
\end{raggedright}

%
\section{Introduction}

The decays $K_L\to\pz\g\g$, $K_L\to\pz\ee\g$ and $K_L\to\pz\pz\g$
can all be used as tests of chiral perturbation theory. In particular,
predictions for the branching ratios of these modes show significant
increases when one uses $O(p^6)$ versus $O(p^4)$ chiral perturbation
theory. The first measurements of $K_L\to\pz\g\g$ 
\cite{na31_pi0gg,e731_pi0gg}\ were a factor of
three higher than the $O(p^4)$ prediction, but were consistent with
the $O(p^6)$ calculation.\cite{dambrosio1} This was also seen in the decay
$K_L\to\pz\ee\g$ decay where the $O(p^4)$ prediction was inconsistent
with the measurement, but consistent with the $O(p^6)$ 
calculation.\cite{donoghue}
For the $K_L\to\pz\pz\g$ decay, the branching ratio vanishes
at $O(p^4)$ chiral perturbation theory, yet is non-zero at higher
order. Recent predictions for this decay range from $10^{-11}-10^{-8}$.
\cite{pzpzg1, pzpzg2}

In addition, the two decays $K_L\to\pz\g\g$ and $K_L\to\pz\ee\g$
are important for understanding the direct CP violating decay
$K_L\to\pz\ee$. Three components contribute to the
$K_L\to\pz\ee$ amplitude:  direct CP violation, indirect
CP violation (and an interference term), and a CP conserving
term. Recent measurements of the decay $K_S\to\pz\ee$\cite{na48_pi0ee}
and $K_S\to\pz\mm$\cite{na48_pi0mm} have helped to determine the
indirect CP violating contributions to $K_L\to\pz\ee$ and $K_L\to\pz\mm$. The
magnitude of the CP conserving contributions to $K_L\to\pz\ll$ can
be determined by measurements of the decay 
$K_L\to\pz\g\g$\cite{na48_pi0gg, ktev_pi0gg} and $K_L\to\pz\ee\g$.
The CP conserving term is estimated to be small\cite{dambrosio}.

\section{The KTeV Experiment}
\begin{figure}[htb]
  \begin{center}
  \epsfig{file=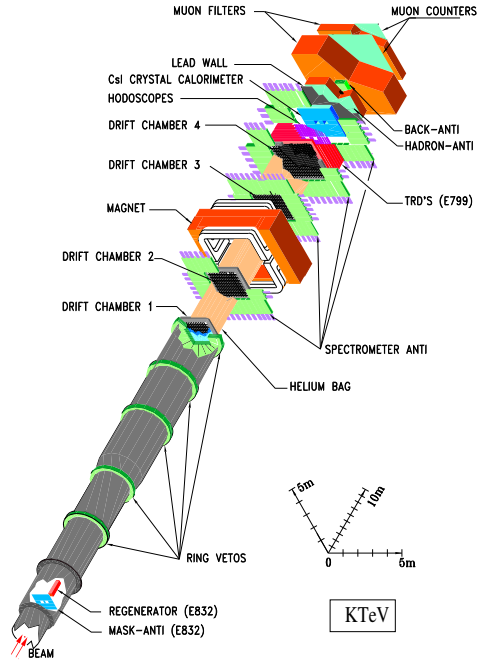,height=9.0cm}
  \caption{ The KTeV detector.
  \label{fig:ktev_det} }
  \end{center}
\end{figure}

The KTeV experiment, shown in Figure~\ref{fig:ktev_det}, is a Fermilab
fixed target experiment. The detector ran in two different configurations:
E799 and E832. The E799 configuration took advantage of a higher kaon flux
to search for rare kaon decays. The E832 configuration was used primarily
for a measurement of $\ep'/\ep$. The main difference between the two
configurations was the use of a regenerator to produce $K_S$ decays
in the E832 configuration. The two experimental configurations also
had a few other differences as noted below.

The KTeV detector contains a charged spectrometer with four drift chambers,
two on either side of a large dipole magnet. At the downstream end
of the detector is a two-meter square calorimeter consisting
of 3100 pure CsI blocks. 
Following the calorimeter are 10 cm of lead
and 5 meters of steel which act as a muon filter.
Two planes of scintillator, used for muon detection, are located
just downstream of the steel.  Photon vetoes to detect the
the presence of particles that would otherwise escape detection
surround the spectrometer. The charged spectrometer
achieves a hit resolution of better than 100 $\mu$m, while the CsI
calorimeter obtains better than 1\% energy resolution over the 
range of energies of interest.
Just upstream of the CsI calorimeter
is a transition radiation detector (TRD) capable of $e/\pi$ separation of
200:1 with a 90\% efficiency. The TRDs were employed during E799 running,
and were moved out of the way during E832 running.

The KTeV experiment took data during a number of different periods between
1996 and 1999. In the E832 configuration, we had three different running
periods: 1996, 1997 and 1999. The E799 running occurred in 1997 and 1999.
Between the 1997 and 1999 runs, a number of upgrades were made
to the detector to increase its reliability and to improve
its live time. In addition during E799 running,
the transverse kick from the magnet was
reduced from 205 MeV/$c$ to 150 MeV/$c$, enabling a larger acceptance
for high multiplicity decay modes.
For the entire E799 data set, approximately $6.6\times 10^{11}$ kaon
decays occurred
in the KTeV detector. This large kaon flux
allows us to have an unprecedented sensitivity to a number of
rare kaon decays with large multiplicity final states.

\begin{figure}[htb]
  \begin{center}
    \epsfig{file=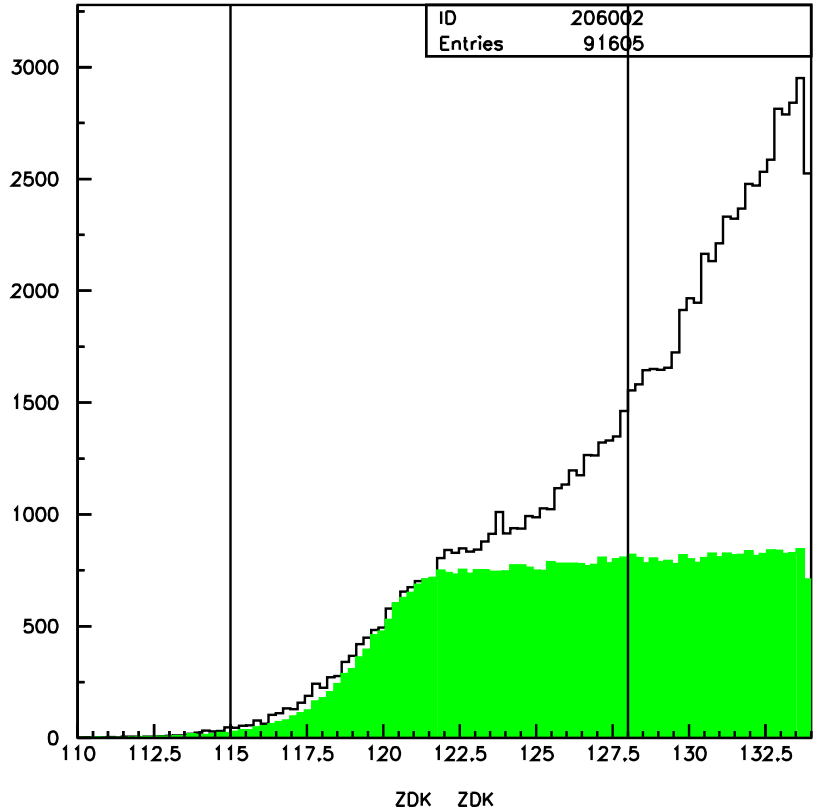,width=0.8\textwidth}
    \caption{ The kaon reconstructed decay position. The green histogram
                 is the $K_L\to\pz\g\g$ Monte Carlo, while the black histogram
                 is the $K_L\to\pz\pz\pz$ Monte Carlo.
    \label{fig:pggzdk} }
  \end{center}
\end{figure}

\begin{figure}[htb]
  \begin{center}
    \epsfig{file=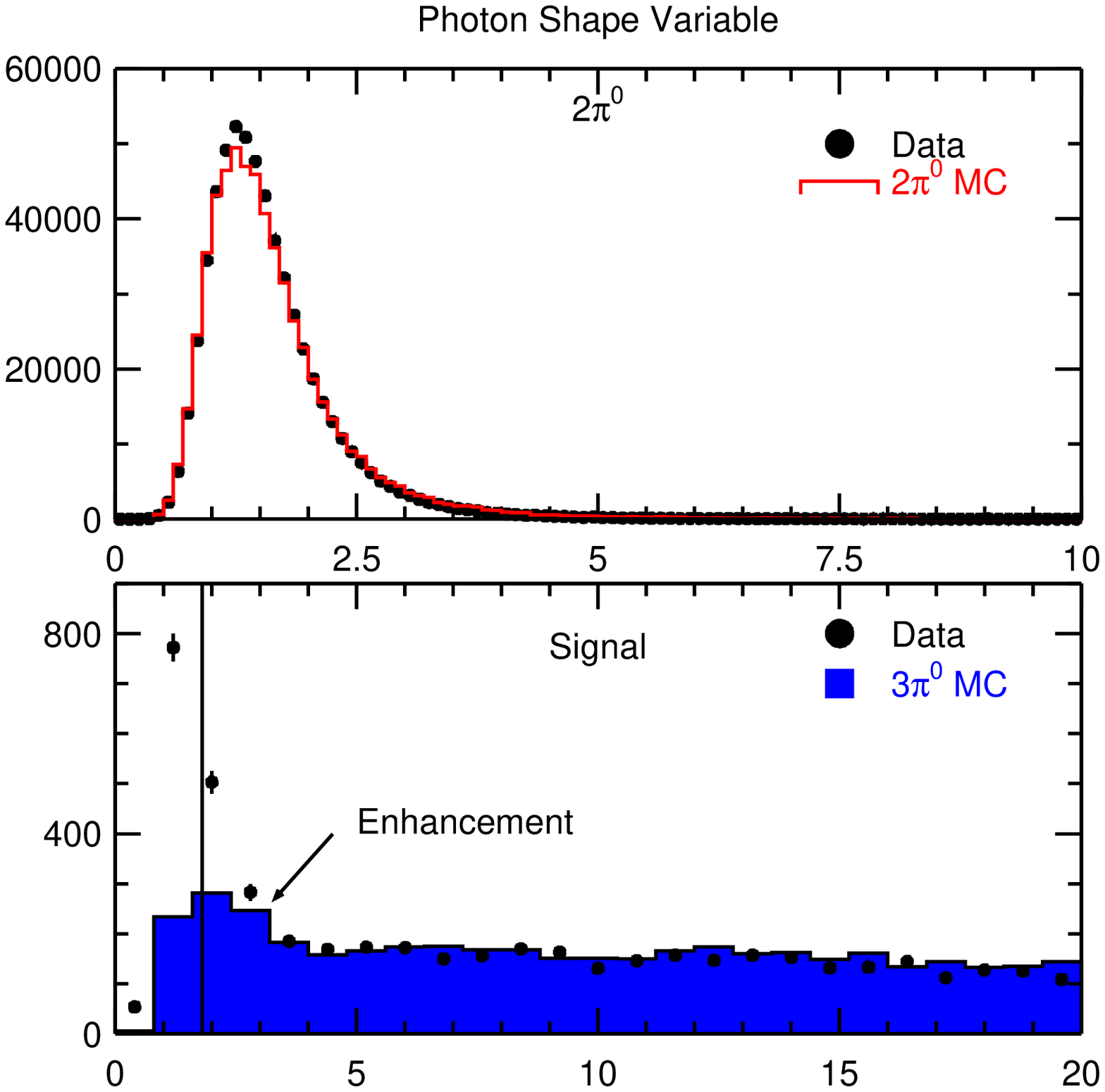,width=0.8\textwidth}
    \caption{ The photon shape variable for $K_L\to\pz\pz$ (top) and
                 $K_L\to\pz\g\g$ candidates (bottom). In the top plot, the
                 dots are the data and the red histogram is the Monte Carlo.
                 In the bottom plot, the blue histogram represents the 
                 $K_L\to\pz\pz\pz$ background.
    }
    \label{fig:fuse3x3_2pi0_dsh}
  \end{center}
\end{figure}

\section{Measurement of $K_L\to\pz\g\g$}
In the decay $K_L\to\pz\g\g$, the final state consists of four photons,
two from the $\pz$ and two from the $K_L$. Our analysis requires
four electromagnetic clusters in the CsI calorimeter, with no
tracks or extra clusters. One can combine the four clusters in 
three different combinations. We choose the combination in which 
the two photon invariant mass reconstructs
closest to the known $\pz$ mass. 

Backgrounds to this decay originate from two main sources, $K_L\to\pz\pz$
and $K_L\to\pz\pz\pz$. The first background has the same topology
as our signal events. However, the relatively small branching ratio
for $K_L\to\pz\pz$ and kinematic cuts reduce this background to
a negligible level. In particular, we look at all possible photon
combinations of the four photons and reject any event in which the
invariant $\g\g$ mass for both pairs of photons
reconstruct near the $\pz$ mass. 

$K_L\to\pz\pz\pz$ decays constitute the largest source of background
to $K_L\to\pz\g\g$. These events can contribute to the background 
in three distinct ways. First, two photons from the $K_L\to\pz\pz\pz$ 
decay can miss the calorimeter, with the remaining four photons interacting
in the CsI calorimeter. Another possibility is that one photon misses
the calorimeter while two photons overlap in the calorimeter. Finally,
all six photons can hit the calorimeter, with 
four of the six photons from the $K_L\to\pz\pz\pz$ decay overlapping in the
calorimeter, producing four separate clusters in the calorimeter.

The first source of $K_L\to\pz\pz\pz$ background can be reduced
by first eliminating events with signals in the photon vetoes.
In addition, one can improve the signal to background ratio
by cutting on 
the reconstructed $z$ position of the event. When the energy in the 
calorimeter is less than the kaon energy, the event reconstructs
downstream of its true decay position. This can be seen in
Figure~\ref{fig:pggzdk} where the signal events are relatively flat,
while the background events show a large enhancement at the 
downstream end of the detector.

To get rid of events in which photons overlap in the CsI calorimeter,
we define a photon shape variable. This variable uses the 3x3
array of CsI crystals containing the core of the shower and compares
the energy distribution to an ideal energy distribution determined
from Monte Carlo. This shape variable can be seen in 
Figure~\ref{fig:fuse3x3_2pi0_dsh}. The top plot shows the photon shape
variable for $K_L\to\pz\pz$ events for both the data and the Monte Carlo,
and shows good agreement between the two.
The bottom plot shows the data with the $K_L\to\pz\pz\pz$ Monte
Carlo overlaid. As can be seen, the signal events tend to peak at
low values of the shape variable, while the $3\pz$ background
is relatively flat. We require that
the shape variable be less than 1.8, which removes the majority
of the $3\pz$ events.

Because of the discrepancy between the NA48 and KTeV published
results on $K_L\to\pz\g\g$, we reexamined the data used 
in our previous result.\cite{ktev_pi0gg} We found that
we underestimated the impact of a disagreement between the
data and the Monte Carlo photon shape variable.
This led to an underestimate of
the background in our final sample. With an improved simulation
of the photon shape variable, our background estimate nearly doubles,
and the $K_L\to\pz\g\g$ branching ratio decreases.

\begin{figure}[htb]
  \begin{center}
    \epsfig{file=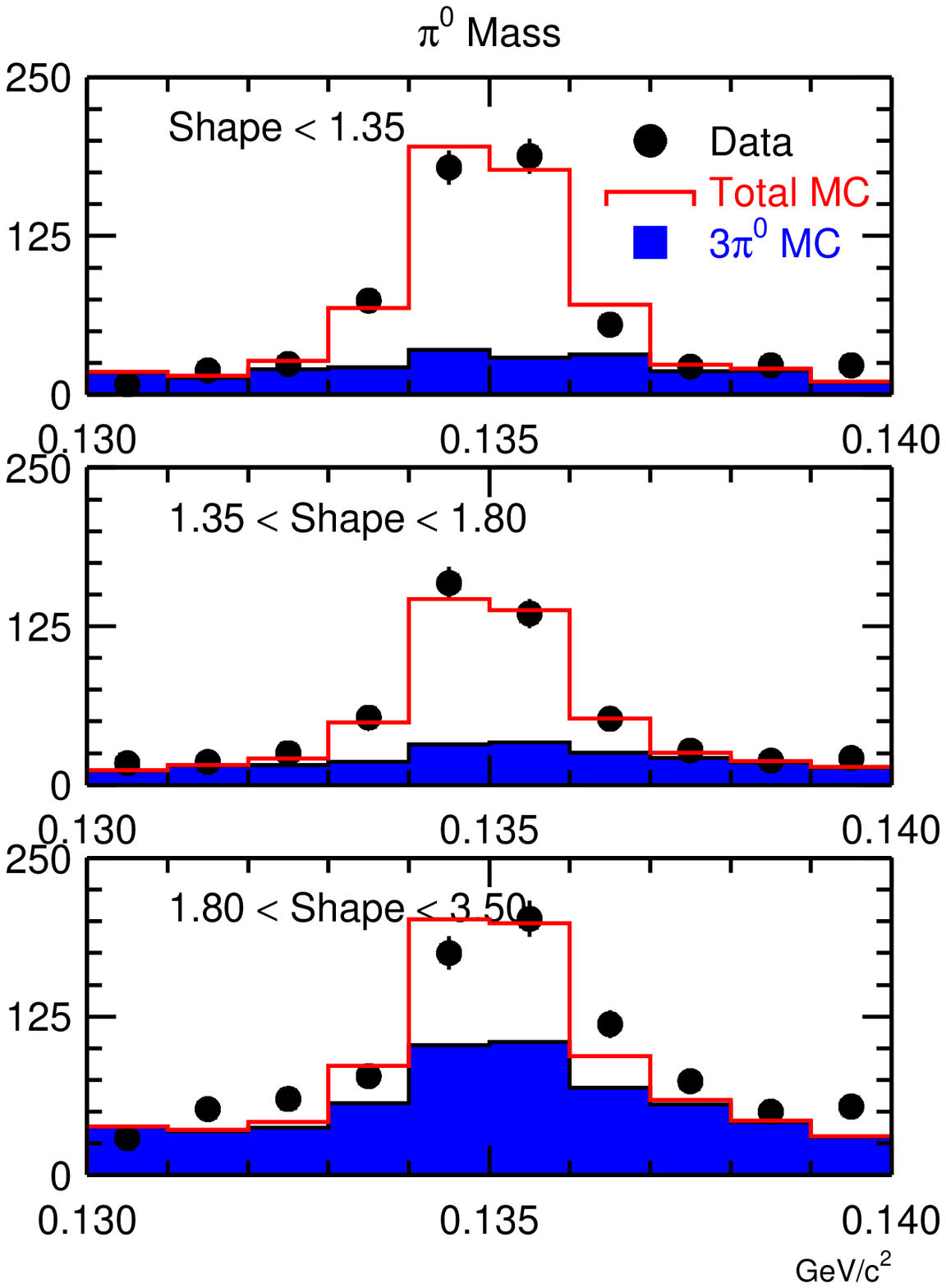,width=\textwidth}
    \caption{ The reconstructed $\pz$ mass distribution for
       candidate events. The dots are the data, the red histogram the
       sum of the Monte Carlo, and the blue histogram the $K_L\to\pz\pz\pz$
       background Monte Carlo. The three plots represent different regions
       of the photon shape variable.
    }
    \label{fig:pi0m_overlays_dsh}
  \end{center}
\end{figure}

After all cuts, the background is dominated by $K_L\to\pz\pz\pz$ events,
with the total background constituting approximately 30\% of the 
final events. To check that we understand the background level,
we examine the data/MC overlays of the $m_{\pz}$ distributions.
Any underestimate of the $3\pz$ background would manifest itself
as a data/MC mismatch in the tails of the $m_{\pz}$ distributions.
As shown in Figure~\ref{fig:pi0m_overlays_dsh}, at small values of
the photon shape variable, the tails agree quite well in the
$m_{\pz}$ distributions.

The systematic errors associated with this measurement are shown
in Table~\ref{tab:pi0gg_syst}. The main sources of uncertainty stem
from understanding the $3\pz$ background and its normalization. Other
sources of systematic uncertainty come from our knowledge of the
photon veto system, the acceptance determination and external factors
such as the measured $K_L\to\pz\pz$ branching ratios. The total
systematic uncertainty is 2.9\%.

\begin{table}[htb]
  \begin{center}
    \begin{tabular}{lll} \hline\hline
    Source of Uncertainty & Uncertainty (\%) \\ \hline
    $a_V$ dependence       & 1.5 & \\ 
    $3\pz$ background      & 1.3 &\\
    \hspace{.5cm} MC statistics & & 1.0 \\
    \hspace{.5cm} Normalization & & 0.9 \\
    Photon Shape           &  1.1 & \\
    Tracking Chambers      & 0.9  & \\
    $2\pz$ branching ratio & 0.9 &\\
    Photon vetoes          & 0.9 & \\
    Kaon Energy            & 0.7  & \\
    Decay Vertex           & 0.4  & \\\hline
    Total                  & 2.9 \\
    \hline \hline
    \end{tabular}
    \label{tab:pi0gg_syst}
    \caption{Systematic errors for the $K_L\to\pz\g\g$ measurement}
  \end{center}
\end{table}

\begin{figure}[htb]
  \begin{center}
    \epsfig{file=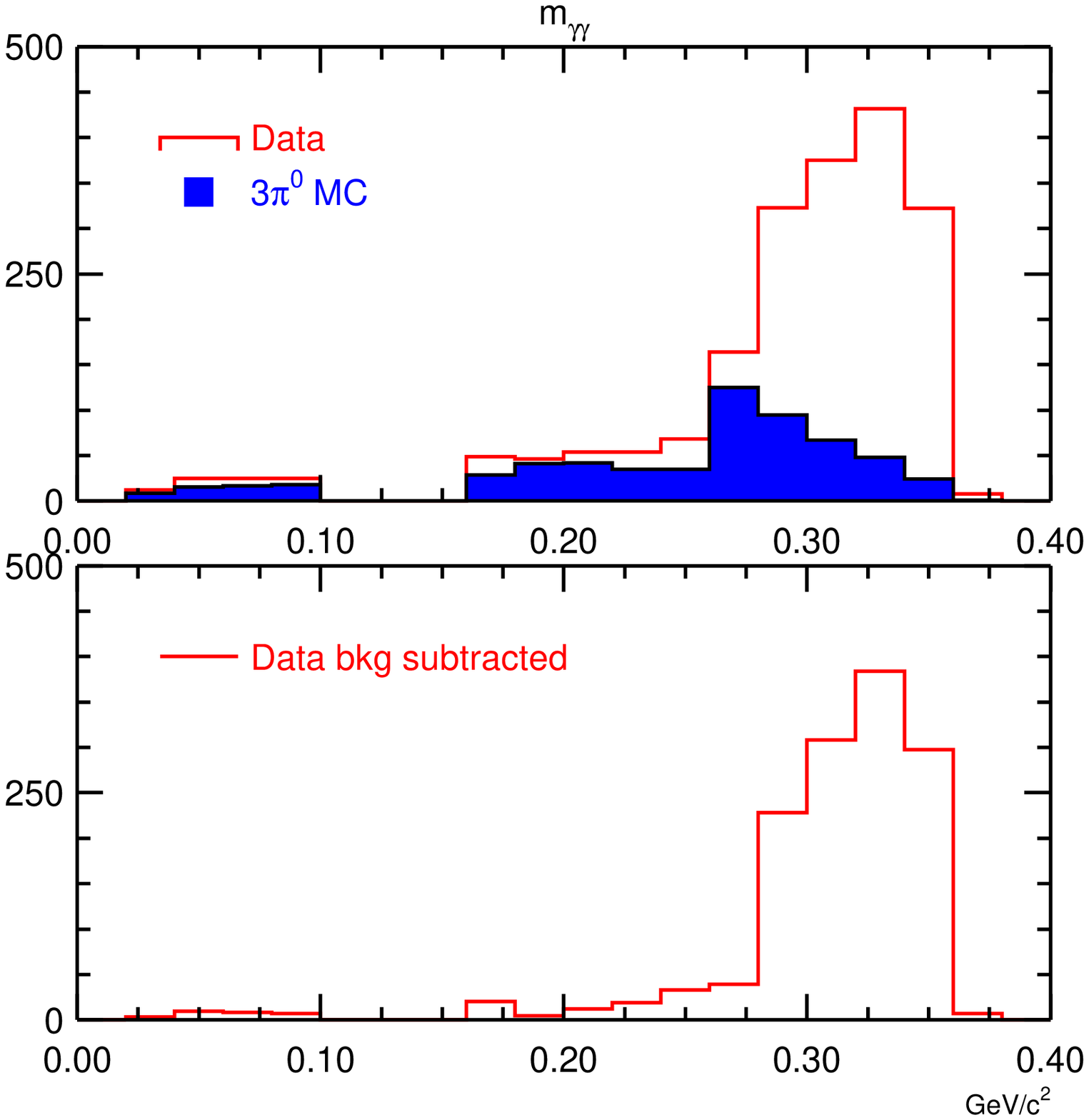,width=0.78\textwidth}

    \epsfig{file=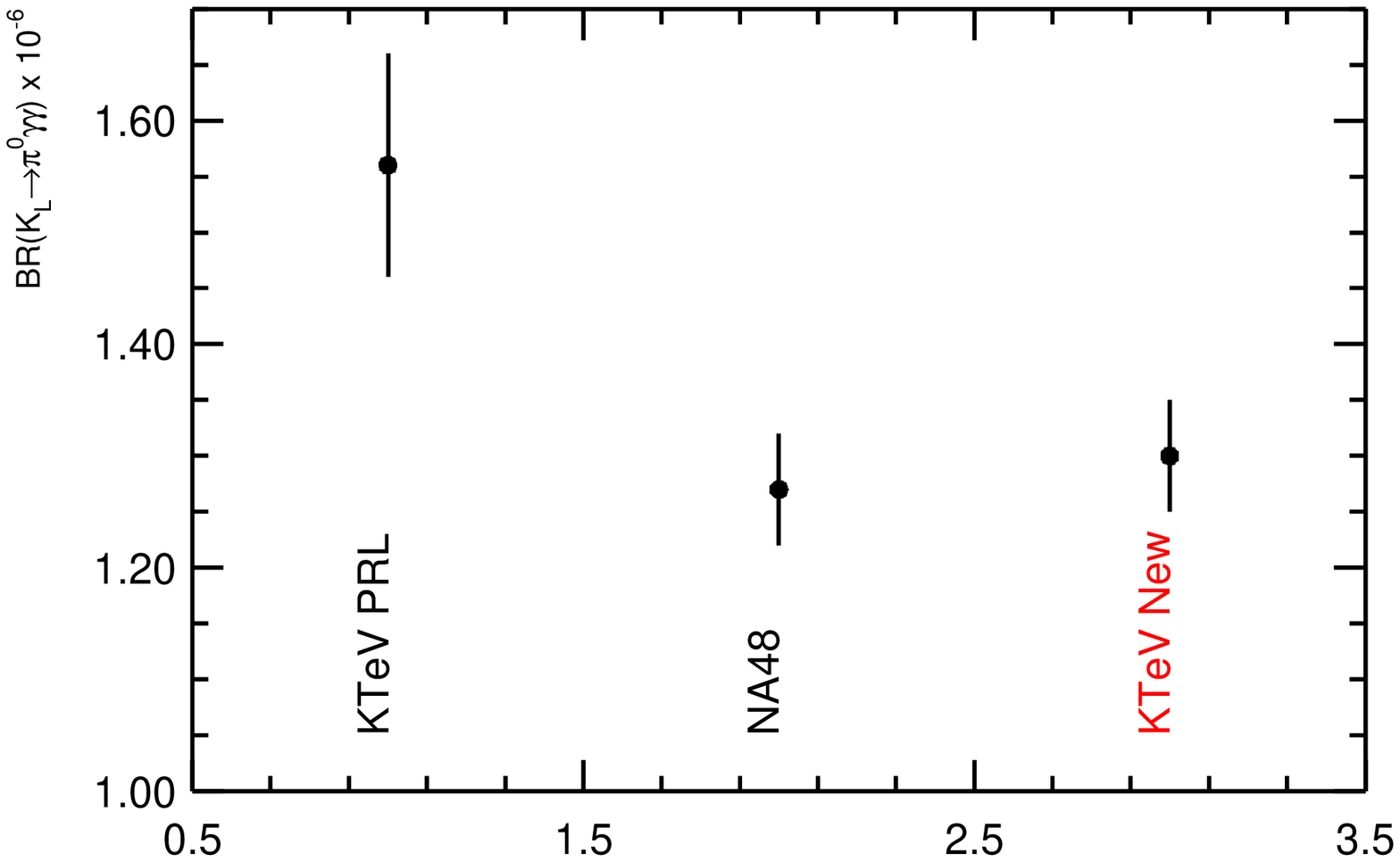,width=0.78\textwidth}
    \caption{ The top plots show the final $\g\g$ mass distribution for
       $K_L\to\pz\g\g$ candidates before and after background subtraction.
       The bottom plot shows the branching ratio results for the
       previous KTeV result, the NA48 result and our new KTeV result.
       The results have been rescaled using the latest PDG value for
       $K_L\to\pz\pz$.
    }
    \label{fig:mgg_all}
  \end{center}
\end{figure}

After all cuts have been implemented, we find 1982 events with a
background of 601 events. The reconstructed $\g\g$ mass is shown
in Figure~\ref{fig:mgg_all}. The distinctive $\g\g$ shape results
from coupling of the $\g\g$ system to two virtual pions, and peaks
around 320 MeV/$c^2$. We determine the $K_L\to\pz\g\g$ branching
ratio to be: BR($K_L\to\pz\g\g$) = $(1.30\pm 0.03 \pm 0.04)\times 10^{-6}$, 
where
the first error is statistical and the second error systematic.
As shown in Figure~\ref{fig:mgg_all}, this result is compatible and
competitive with the result from NA48. Our new result supercedes 
the previous KTeV measurement of BR($K_L\to\pz\g\g$), which was
in nominal disagreement with the NA48 result.

%
%
\section{Measurement of $K_L\to\pz\ee\g$}
The decay $K_L\to\pz\ee\g$ produces two charged tracks in the
spectrometer and three photons in the CsI calorimeter. The three
photons can be combined in three different ways to form a neutral
pion. We choose the combination that has the best $\pz$ mass. The
neutral vertex is used to determine the decay position rather than
the charged vertex due to its better resolution.

\begin{figure}[htb]
  \begin{center}
    \epsfig{file=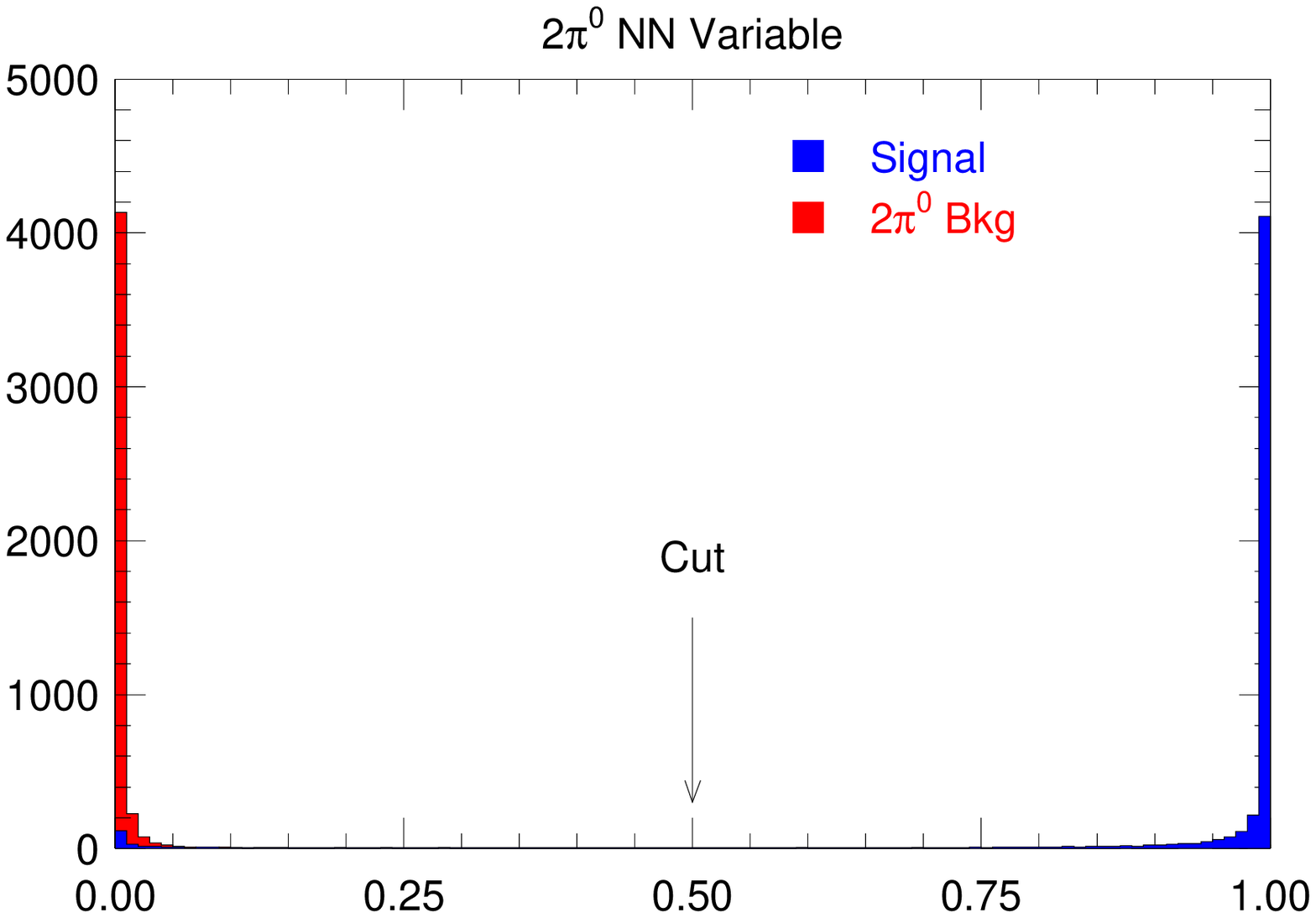,width=0.75\textwidth}

    \epsfig{file=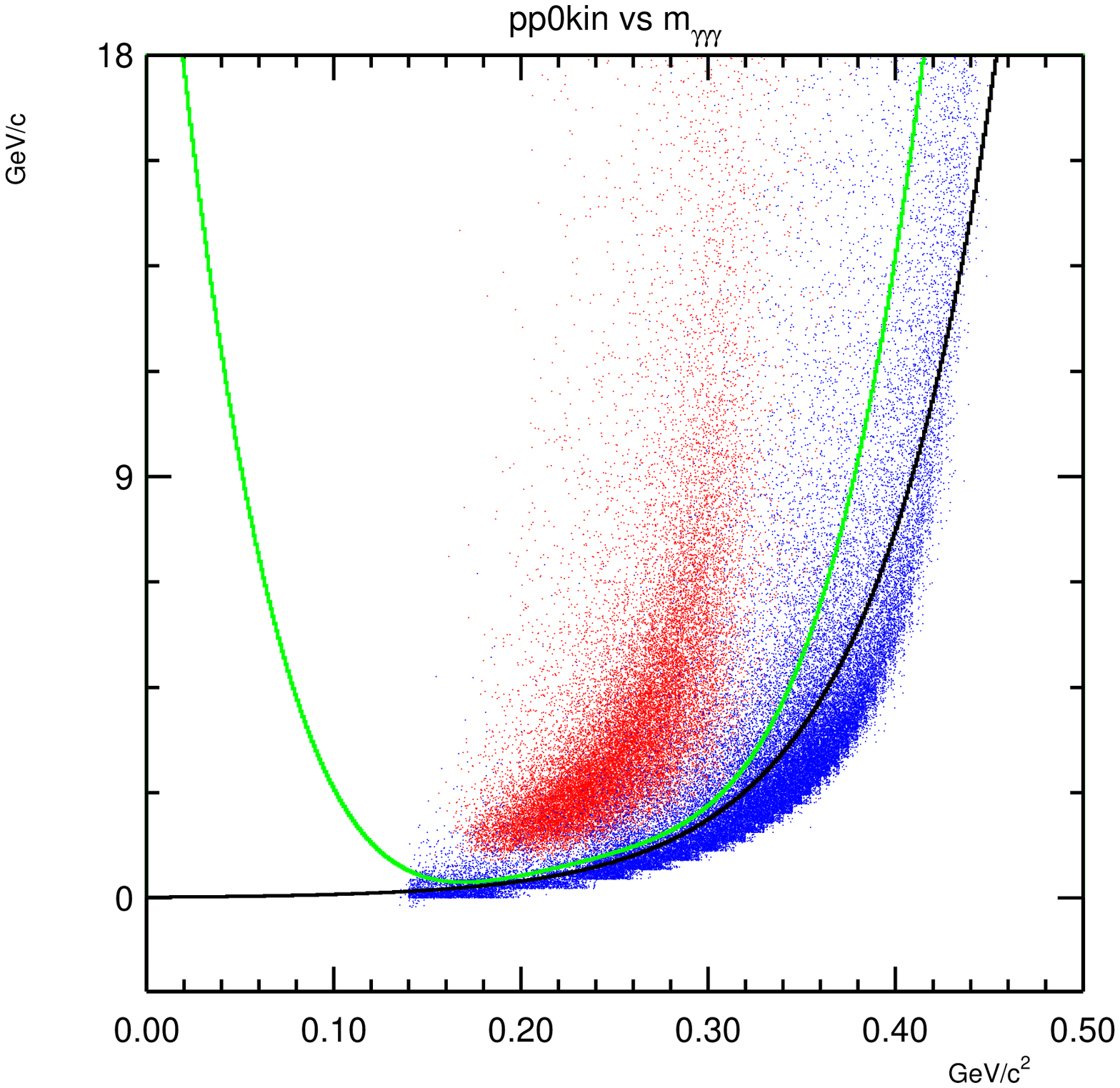,width=0.75\textwidth}
    \caption{ The variables used to reduce the $K_L\to\pz\pz$ (top) and
      $K_L\to\pz\pz\pz$ (bottom) backgrounds. The top plot shows the
      neural net variable for the $K_L\to\pz\pz$ background (red) and
      the $K_L\to\pz\ee\g$ signal Monte Carlo. The bottom plot shows the
      missing momentum versus the $\g\g\g$ mass distribution for
      $K_L\to\pz\pz\pz$ background (red) and signal Monte Carlo (blue).
    }
    \label{fig:2pi0NN}
  \end{center}
\end{figure}

Like the $K_L\to\pz\g\g$ decay, the main backgrounds to this decay
are from $K_L\to\pz\pz$ and $K_L\to\pz\pz\pz$. The difference is that
one of the $\pz$ undergoes Dalitz decay to $\ee\g$. To help reduce
the background from $K_L\to\pz\pz$ decays, we formed a neural net
variable using the reconstructed masses $m_{\g\g}$  and $m_{\ee\g}$
from the second and third best combinations. We define the second
and third best combinations by how far from the nominal $\pz$
mass the ${\g\g}$ combination reconstructs. The neural net variable can be seen
in Figure~\ref{fig:2pi0NN}. In this plot the $K_L\to\pz\pz$ events
are well-separated from the signal $K_L\to\pz\ee\g$ events. We
require that the neural net variable be greater than 0.5

To reduce $K_L\to\pz\pz\pz$ backgrounds we also require that the
photon shape variable, defined above, be small for each photon
candidate. We can use the decay kinematics to help reduce the
background from $K_L\to\pz\pz\pz$ decays. We define a variable
$p^2_L$, which is the longitudinal momentum squared of the missing
$\pz$ in the kaon rest frame.
We perform a two-dimensional cut on the 
$p^2_L$ variable versus the three 
photon invariant mass, $m_{\g\g\g}$. This distribution is shown in 
Figure~\ref{fig:2pi0NN}. There is good separation between the
signal events, shown in blue, and the $K_L\to\pz\pz\pz$ background.
We use a polynomial to define the cut shown in Figure~\ref{fig:2pi0NN}.

\begin{figure}[htb]
  \begin{center}
    \epsfig{file=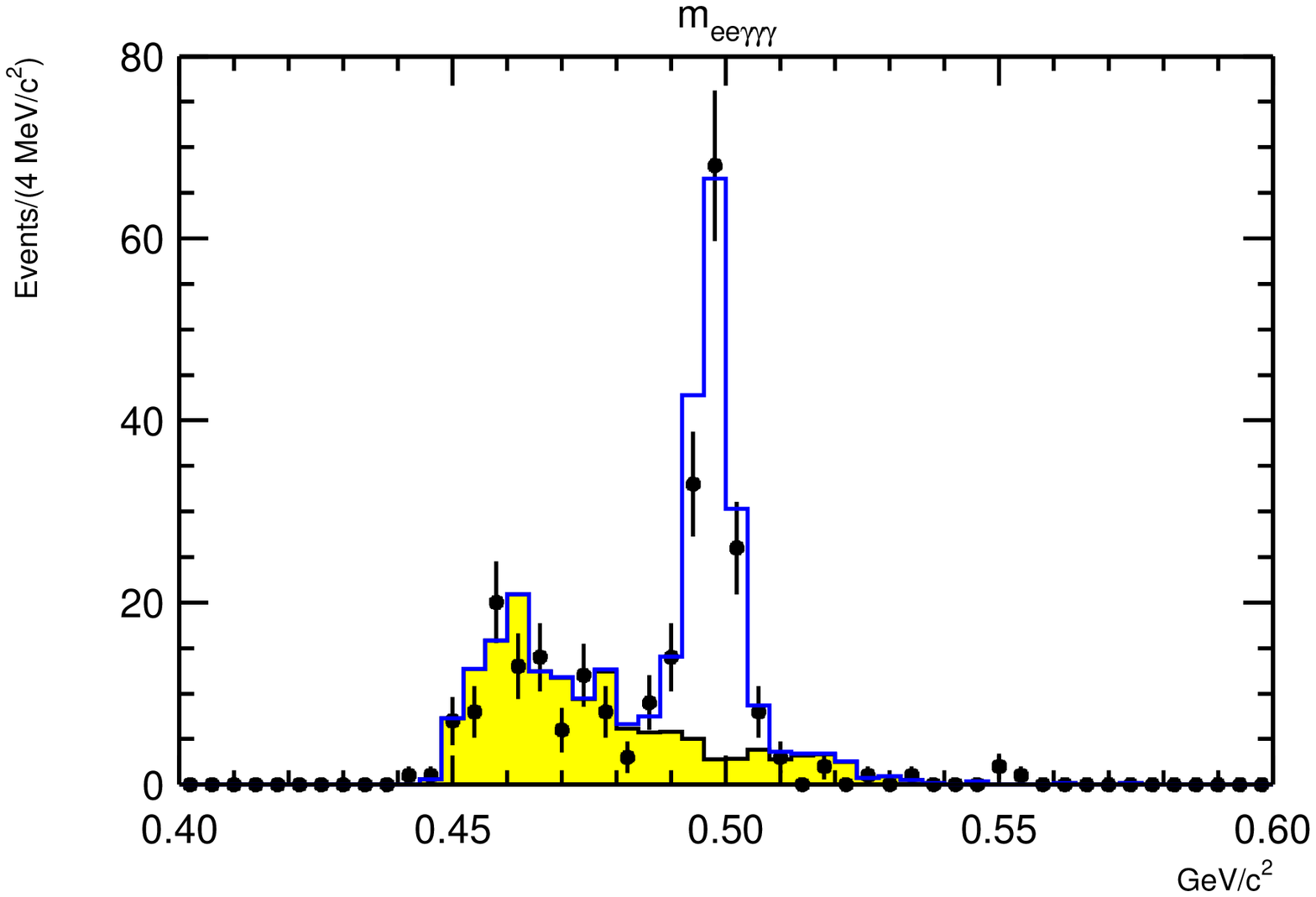,width=0.7\textwidth}

    \epsfig{file=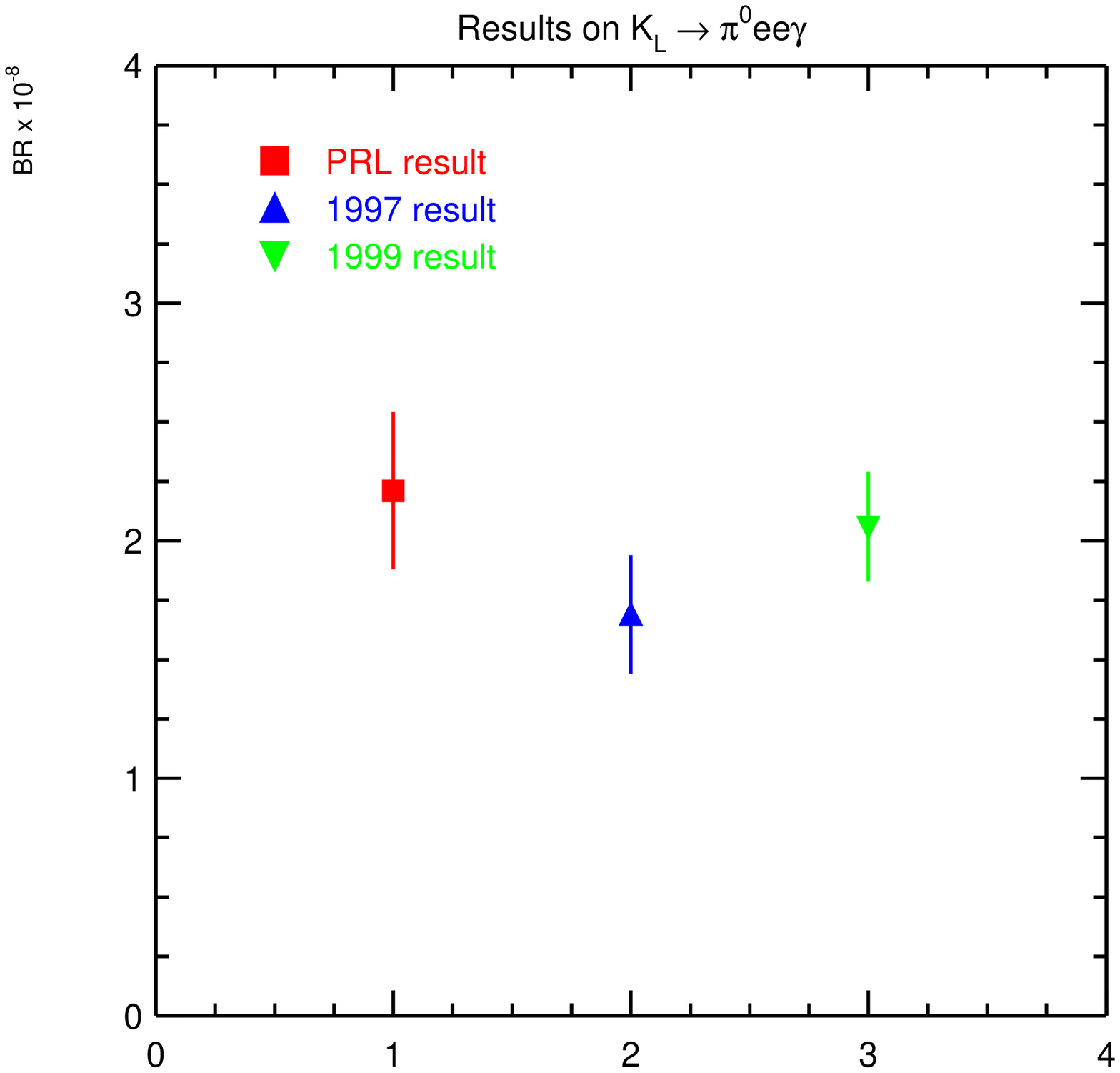,width=0.7\textwidth}
    \caption{The $\ee\g\g\g$ mass (top) for the combined 1997 and 
      1999 data sets. The dots are the data,
      the blue histogram is the sum of the signal plus background Monte
      Carlo, and the yellow histogram represents the sum of the
      $K_L\to\pz\pz$ and $K_L\to\pz\pz\pz$ backgrounds. The bottom plot
      shows the branching ratio results for the KTeV PRL (red)
      and our new 1997 (blue) and 1999 (green) analyses.
      The results have been rescaled using the latest PDG value for
      $K_L\to\pz\pz$.
   }
   \label{fig:meeggg_final}
  \end{center}
\end{figure}

After applying all cuts, we find 139 events over a background of
14.4 events as shown in 
Figure~\ref{fig:meeggg_final}. We reconstructed 80,445 $K_L\to\pz\pz$ events
used for normalization. This allows us to determine the branching ratio
for $K_L\to\pz\ee\g$ to be
BR($K_L\to\pz\ee\g$) = $(1.90\pm 0.16\pm0.12)\times 10^{-8}$
where the first error is statistical and the second
is systematic. The results from the 1997 and 1999 data sets
are shown separately in Figure~\ref{fig:meeggg_final}, along with
the previously published result on $K_L\to\pz\ee\g$.  The published
result and our new 1997 result use the same data set. Although
the branching ratio results for these two results differ, the
two results are statistically consistent with each other. We 
expect the two results to differ because of 
differences in selection criteria and calibration constants.
As in the $K_L\to\pz\g\g$, the normalization mode
branching ratio has decreased by approximately 8\% from the value
used in the published KTeV result. The systematic uncertainties
are listed in Table~\ref{tab:pi0eeg_syst}. The largest systematics
come from the limited Monte Carlo statistics and the dependence
of the result on the parameter $a_V$.

\begin{table}[tbh]
  \begin{center}
    \begin{tabular}{l c}
      Systematic & Error (\%)\\ \hline \hline
      MC Statistics          &  4.2   \\
      $a_V$ dependence       &  3.8   \\
      $K_L$ and $\pz$ BR     &  2.8  \\
      $3\pz$ bkg             &  0.8   \\
      acceptance             &  0.4   \\ 
      $2\pz$ background      &  0.1   \\ \hline
      Total                  &  6.4   \\ \hline
      \hline
    \end{tabular}
    \label{tab:pi0eeg_syst}
    \caption{Systematic errors for the $K_L\to\pz\ee\g$ measurement}
  \end{center}
\end{table}

%
%
\section{Search for $K_L\to\pz\pz\g$}
In this decay we use the $\pz$ Dalitz decay for one of the
neutral pions. We chose this because the fully neutral mode trigger
was heavily prescaled, whereas the KTeV two-track trigger was not.
The final state consists of an $\ee$ pair,
and three photons in the CsI calorimeter. 
As in the other analyses discussed in this
paper, there are three possible combinations of photons for each
event. Again, we choose the combination that reconstructs closest
to the $\pz$ mass. 

The main background to this decay comes from $K_L\to\pz\pz\pz$ events,
since $K_L\to\pz\pz$ cannot contribute to the background without the
addition of accidental particles.
Two variables are effective in
reducing the background to this mode. The first is to employ the
photon shape variable described above. This reduces events in which
two of the photons overlap in the calorimeter. The second varible
used is the missing momentum in the kaon rest frame, $p_L^2$.
Cutting on this variable significantly reduces the
$K_L\to\pz\pz\pz$ background. 

A unique feature of this analysis is that the normalization mode
is not fully reconstructed. Rather we use $K_L\to\pz\pz\pz$ events
in which one of the photons passes through one of the beam holes
in the CsI calorimeter. This photon is kinematically constrained 
using the known kaon mass.

\begin{figure}[htb]
  \begin{center}
    \epsfig{file=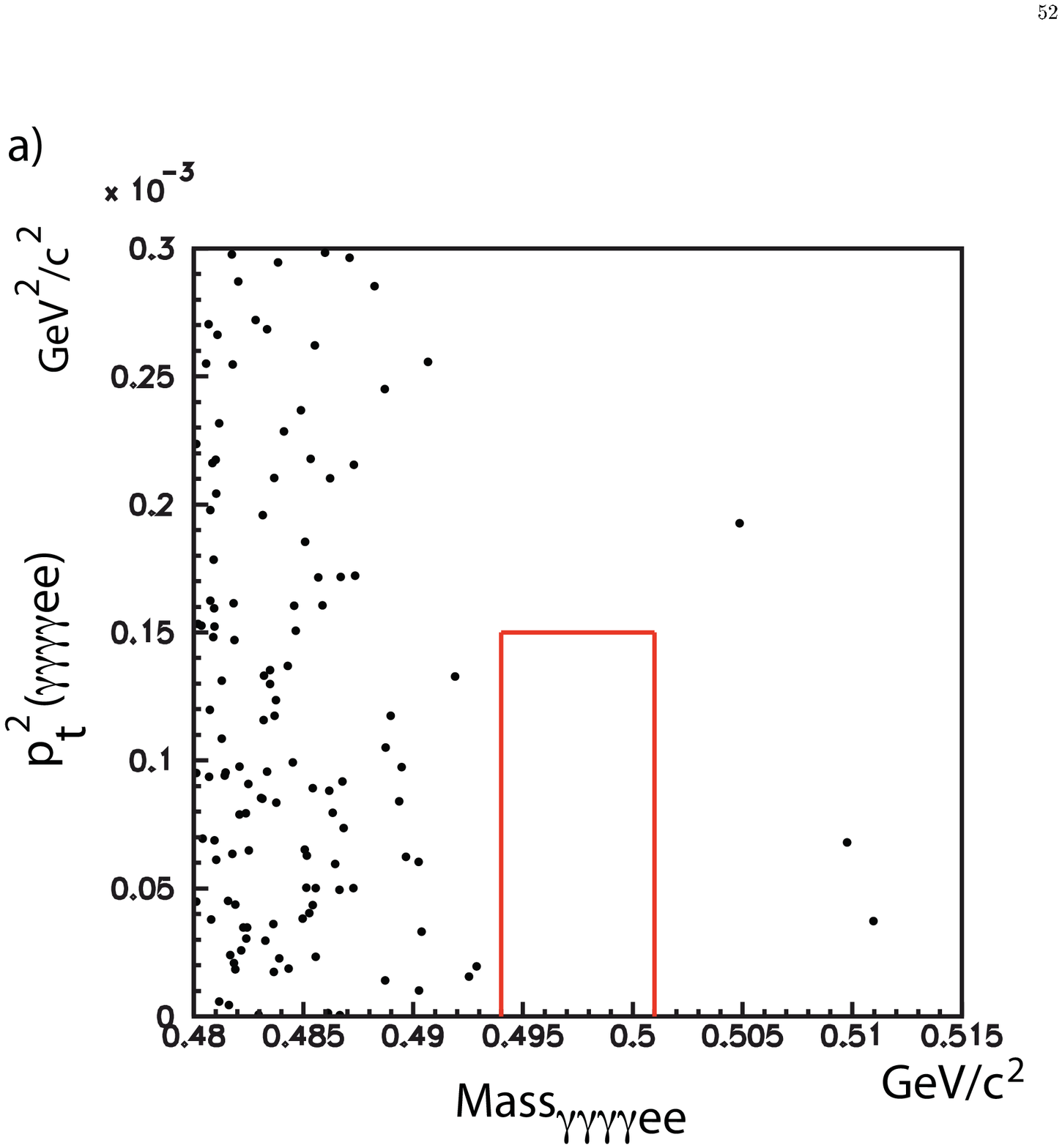,width=0.61\textwidth}

    \epsfig{file=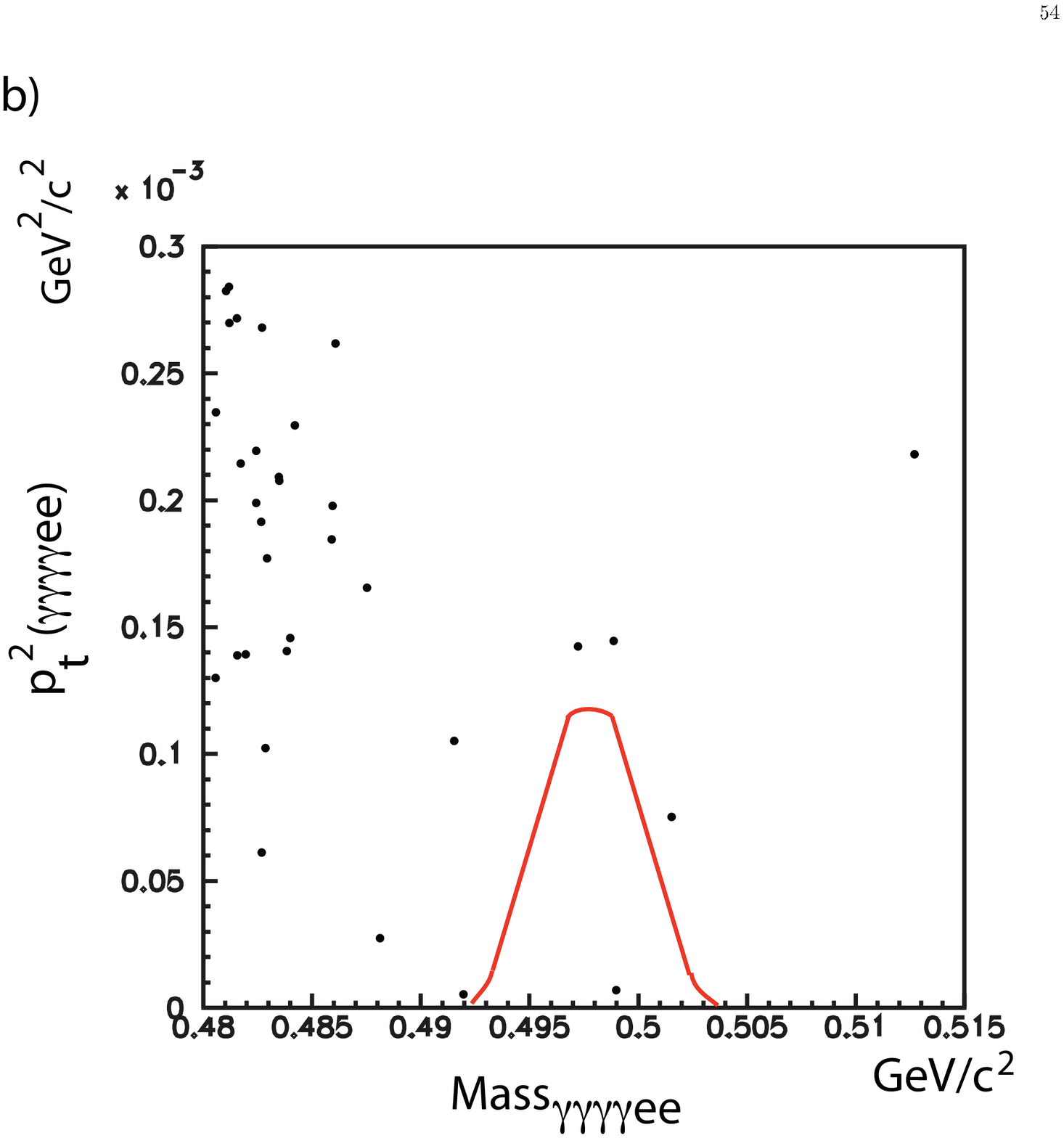,width=0.61\textwidth}
    \caption{The $p^2_T$ versus kaon mass distributions for the 1997 (top)
      and 1999 (bottom) samples after all selection cuts have been applied.
      The red line represents the signal region.
    }
    \label{fig:data_box97}
  \end{center}
\end{figure}

The final candidates are shown in Figures~\ref{fig:data_box97},
where the reconstructed kaon mass is shown
along the $x$-axis and the kaon transverse momentum squared, $p_T^2$,
is shown on the $y$-axis. For the 1997 sample we chose a square
signal region, while for the 1999 sample we formed a likelihood
from the product of the $m_K$ and $p_T^2$ probabilities. After all
cuts we found no events in the 1997 sample and one event in the
1999 sample. The probability of this event to come from background
is determined to be approximately 10\%. 

Using our $K_L\to\pz\pz\pz$ normalization sample, we determine an
upper limit of BR($K_L\to\pz\pz\g$) $< 2.32\times 10^{-7}$. This
constitutes an improvement of about 20 over the previous 
limit from NA31. However, this limit is well-above the predictions
for this decay.

\section{Summary and Conclusions}
The KTeV experiment has presented three new results on 
$K_L\to\pz\g\g$, $K_L\to\pz\ee\g$ and $K_L\to\pz\pz\g$. The first
result is competitive with the world's best result from NA48,
while the other two represent the world's best measurements
on these two decays. The measured branching ratios are both
inconsistent with the $O(p^4)$ predictions from chiral perturbation
theory, and consistent with $O(p^6)$ chiral perturbation
theory.


\begin{thebibliography}{99}
\bibitem{na31_pi0gg} G.D. Barr {\it et al.}, Phys. Lett. {\bf B284}, 
  440 (1992).
\bibitem{e731_pi0gg} V. Papadimitriou {\it et al.} Phys. Rev. {\bf D44},
  573 (1991).
\bibitem{dambrosio1}G. D'Ambrosio and J. Portoles, Nucl. Phys. {\bf B492},
  417 (1997).
\bibitem{donoghue} J. Donoghue and F. Gabbiani, 
Phys. Rev. {\bf D56}, 1605 (1997).
\bibitem{pzpzg1} P. Heiliger and L.M. Sehgal, Phys. Lett. {\bf B307}, 
  182 (1993).
\bibitem{pzpzg2} G. Ecker, H. Neufeld and A. Pich, 
    Nucl. Phys. {\bf B413}, 321 (1994).
\bibitem{na48_pi0ee} J.R. Batley {\it et al.}, Phys. Lett. {\bf B576}, 
  43 (2003).
\bibitem{na48_pi0mm} J.R. Batley {\it et al.}, Phys. Lett. {\bf B599},
  197 (2004)..
\bibitem{na48_pi0gg} A. Lai {\it et al.}, Phys. Lett. {\bf B536}, 229 (2002).
\bibitem{ktev_pi0gg} A. Alavi-Harati {\it et al.}, Phys. Rev. Lett. {\bf83},
  917 (1999).
\bibitem{dambrosio}G. D'Ambrosio, G. Ecker,
  G. Isidori and J. Portoles, J. High Energy Physics,
  {\bf 08}, 004 (1998).
\end{thebibliography}
\end{document}